\documentclass[prd,twocolumn,floats,superscriptaddress,eqsecnum,floatfix]{
revtex4}

\usepackage{amssymb,amsmath}
\usepackage{epsfig}

\newcommand{\be}{\begin{equation}}
\newcommand{\ee}{\end{equation}}
\newcommand{\ba}{\begin{eqnarray}}
\newcommand{\ea}{\end{eqnarray}}

\newcommand{\const}{\mbox{const}}

\newcommand{\eq}[1]{Eq.(\ref{#1})}
\newcommand{\n}[1]{\label{#1}}

\newcommand{\hh}{\, ,\hspace{0.8cm}}

\newcommand{\ins}[1]{{\mbox{\tiny #1}}}

\begin{document}

Alberta-Thy 10-12\\
\\

\title{Classical self-energy and anomaly}

\author{Valeri P. Frolov}%
\email[]{vfrolov@ualberta.ca}
\affiliation{Theoretical Physics Institute, Department of
Physics,
University of Alberta,\\
Edmonton, Alberta, Canada T6G 2E1
}
\author{Andrei Zelnikov}%
\email[]{zelnikov@ualberta.ca}
\affiliation{Theoretical Physics Institute, Department of Physics,
University of Alberta,\\
Edmonton, Alberta, Canada T6G 2E1
}

%\today

\begin{abstract}

We study the problem of self-energy of pointlike charges in higher
dimensional static spacetimes. Their energy, as a functional of the
spacetime metric, is invariant under a specific continuous
transformation of the metric. We show that the procedure of
regularization of this formally divergent functional breaks this
symmetry and results in an anomalous contribution to the finite
renormalized self-energy. We proposed a method of calculation of this
anomaly and presented an explicit expressions for it in the case of a
scalar charge in four and five-dimensional static spacetimes. 
This anomalous correction proves to be zero in even dimensions, but it
does not vanish in odd-dimensional spacetimes.
\end{abstract}


\maketitle

\section{Introduction}

The problem of calculation of the self-energy of charged particles has a long
history going to classical works
\cite{Thomson:1881,Lorentz:1899,Lorentz:1904,Abraham:1903,Poincare:1905,
Fermi:1921} and many others, where the electromagnetic origin of
the electron mass had been studied. Achievements of renormalization techniques
in quantum field theory contributed a lot to the understanding of the problem.
Classical self-energy of an electron can be derived as the limit
of its quantum value. It had been shown
\cite{Vilenkin_Fomin:1974,Vilenkin:1979}, that there exists a correct
quantum-to-classical correspondence for the self-energy of the electron
(see also \cite{EfimovIvanovMogilevsky:1977}.)

In quantum electrodynamics the self-energy of an electron diverges and,
hence, should be regularized and renormalized. A classical
self-energy of pointlike charges suffers similar divergences.
The simplest way to regularize the energy of a classical charged particle is to smear
the charge distribution. If a size $\varepsilon$ of the smearing
tends to zero, its energy diverges as $\varepsilon^{-1}$, so that after
subtraction of this leading term in the expansion over $\varepsilon$,
one obtains a finite expression for the self energy. The problem is
that the result depends on the details of smearing. 

Using methods similar to those adopted in quantum field theory
allows one to formulate the renormalization procedure for a classical 
charged particle in a more general and efficient form. This occurs
because when the size $\varepsilon$ of a charge becomes smaller than
the corresponding invariant cut-off length, the details of the charge
distribution become unimportant. It should be emphasized that in the 
higher dimensional theories, which are widely discussed in modern
physics, the classical self-energy divergence is much stronger 
$\sim \varepsilon^{-(D-3)}$, where $D$ is the number of spacetime
dimensions. As a result the problem of dependence of the
self-energy 
on the details of the charge distribution becomes more involved. However, 
the covariant renormalization approach adapted from  the quantum field theory 
efficiently cures this `decease' in any number of dimensions (see e.g.
\cite{FrolovZelnikov:2012a}).

In the presence of an external gravitational field the self-energy of
a charged classical particle depends on its position. As a result
there
may exist a non-trivial additional force acting on the particle. This
effect
was discussed in detail for a special case when a charged particle is
located near a static 4-dimensional black hole (see e.g.
\cite{SmithWill:1980,FrolovZelnikov:1982,Lohiya:1982,QuinnWald:1997,Quinn:2000,Poisson:2004, 
ChoTsokarosWisseman:2007}.

In our recent paper \cite{FrolovZelnikov:2012a} we obtained the
expression for the gravitational shift
of the renormalized proper mass of a classical charged particle in the higher-dimensional
Majumdar-Papapetrou spacetimes. We demonstrated how the calculation of
the self-energy of a
pointlike static scalar charge can be reduced to the calculation of the
regularized Green
function of a particular operator defined in the spatial section of static
spacetime. 

We propose to use standard renormalization techniques of quantum
field theory to single out UV divergences. For the problem in question the
point-splitting regularization is particularly convenient, though one can
expect that the other methods like $\zeta-$function regularization, dimensional
regularization, and
others would lead to the same finite answer (see e.g. \cite{Moretti:1999}).
In  the limit of pointlike charges after subtraction of divergences the
dependence of the renormalized self-energy on details of internal structure of
the source disappear and the predictions for the finite self-energy and
self-force become universal.

The proposed approach is well adapted to higher dimensions and leads to
unambiguous universal predictions. In our previous paper
\cite{FrolovZelnikov:2012a} we derived the exact formula for the self-energy
of static scalar charges in higher-dimensional
Majumdar-Papapetrou spacetime \cite{Myers:1986}, which describes a set of
extremally charged black holes in equilibrium. An unexpected property of the
obtained result is
that dimensionality matters. In odd-dimensional spacetimes there appears an
anomalous
contribution to the classical self-energy of charges which depends on the space
curvature. For even spacetime dimensions this anomaly of the self-energy
vanishes, just like trace anomaly vanishes for conformal fields in odd
dimensions. The difference is that the trace anomaly is a quantum effect, while
the self-energy is classical.

The anomaly is a violation of some symmetry. In our case the
classical self-energy of a charge distribution, before taking the
limit of pointlike charge and renormalization of divergences, remains
invariant under continuous symmetry transformations of the metric and the field.
These
local transformations involve simultaneous multiplication of the time
component $g_{tt}$
of the metric on a scalar function of spatial coordinates and conformal
transformation of the spatial part $g_{ab}$ of the spacetime metric.
In the next section we establish the symmetry of the energy functional under
such transformation. The self-energy of pointlike charges formally diverges.
Its regularization and renormalization does not respect this symmetry that
results in the appearance of local anomalous contributions.

%%%%%%%%%%%%%%%%%%%%%%%%%%%%%%%%%%

\section{Self-energy of a scalar charge in a static
spacetime}\label{Self-energy}

Let us consider a minimally coupled massless scalar field $\varPhi$
in a static $D-$dimensional spacetime with the metric $g_{\mu\nu}$
\be\begin{split}\label{dS}
&ds^2=-\alpha^2 dt^2+g_{ab}\,dx^a dx^b \,.
\end{split}\ee
We assume that the spacetime is static $\partial_t\alpha=\partial_t g_{ab}=0$.
The action
for this field is
\be
I=-{1\over 8\pi}\int d^Dy\,\sqrt{-{\rm g}}\,\varPhi^{;\mu}\varPhi_{;\mu}\,.
\ee
Here
\be
{\rm g}=\det g_{\mu\nu}=-\alpha^2\,g\hh
g=\det g_{ab}\, .
\ee
The field obeys the equation
\be\label{BoxPhi}
\Box\,\varPhi=0\, .
\ee

The energy $E$ of a static configuration of fields is
\be\begin{split}
E&={1\over 8\pi}\int d^{D-1}x\,\sqrt{g}\,\alpha\,\varPhi^{;a}\varPhi_{;a}
\, .
\end{split}\ee
It is useful to introduce another field variable $\varphi$
\be
\varPhi=\alpha^{-1/2}\,\varphi\,.
\ee
In terms of this field the energy takes the form
\be\begin{split}\label{E}
E&={1\over 8\pi}\int
d^{D-1}x\,\sqrt{g}\,g^{ab}\left(\varphi_{,a}-{\alpha_{,a}\over
2\alpha}\varphi\right)\left(\varphi_{,b}-{\alpha_{,b}
\over
2\alpha}\varphi\right)
\, .
\end{split}\ee
This expression for the energy formally looks like the Euclidean action of
$(D-1)-$dimensional
scalar field interacting with the external dilaton  field $\alpha$.
One can use this  analogy to reformulate the problem of calculation of the
self-energy in terms of the Euclidean quantum field theory defined on
$(D-1)-$dimensional space and in the presence of the external dilaton field.

The field $\varphi$ satisfies the equation
\be\begin{split}\n{F}
F\,\varphi&\equiv (\triangle  +V)\,\varphi=0\,,\\
V&={(\nabla\alpha)^2\over
4\alpha^2}-{\triangle\alpha\over 2\alpha}\equiv -{\triangle(\alpha^{1/2})\over
\alpha^{1/2}} \, .
\end{split}\ee
Here
\be
\triangle=g^{ab}\nabla_a\nabla_b\,.
\ee

Consider the following transformations of the metric \eq{dS} and the field
$\varphi$
\be\begin{split}\label{trans}
&g_{ab}=\Omega^2 \bar{g}_{ab}\hh \alpha=\Omega^{-n}\bar{\alpha}\,,\\
&\varphi=\Omega^{-n/2}\bar{\varphi}\hh n=D-3\,.
\end{split}\ee
From the point of view of a field theory on  $(D-1)-$dimensional spatial
slice it describes simultaneous conformal transformation of the metric
$g_{ab}$ and transformation of
the  dilaton field $\alpha$. Under these transformations the energy
functional \eq{E}
remains invariant. The operator $F$ in \eq{F} transforms homogeneously
\be\label{barF}
F=\Omega^{-2-{n\over 2}}\,\bar{F}\,\Omega^{n\over 2}\,,
\ee
so that the field equation is invariant under these transformations
\be
(\triangle  +V)\,\varphi = \Omega^{-2-{n\over 2}}(\bar{\triangle}
+\bar{V})\,\bar{\varphi}=0\,.
\ee

The energy $E$ \eq{E} is a functional of $(D-1)-$dimensional dynamical field
$\varphi$ and two external fields $g_{ab}$ and $\alpha$. The transformation
\eq{trans} preserve the value of this functional. In other words our effective
$(D-1)-$dimensional Euclidean field theory is invariant under
infinite-dimensional group parametrized by one function $\Omega(x)$. This
transformation conformally modifies $(D-1)-$dimensional metric $g_{ab}$.
In order to keep $E$  invariant it is necessary to accompany it by additional
transformation of the field $\varphi$ and the dilaton $\alpha$. Let us
emphasize that the equation \eq{BoxPhi} for the minimally coupled scalar field
$\varPhi$ is not conformally invariant.

For point-like sources energy $E$ diverges. To deal
with this divergence one has to use some regularization and renormalization
schemes. The regularized self-energy may not respect the invariance property in
question. In quantum field theory the fact that renormalization procedure breaks
some symmetries of the classical theory is the cause of appearance of conformal,
chiral, etc. anomalies. In our case  the same arguments are applicable to the
renormalized self-energy of classical sources and their self-energy acquires
anomalous terms.
It has been shown \cite{FrolovZelnikov:2012a} that the self-energy of
point-like charges can be written in the form
\be\label{E_ren}
E_\ins{ren}=\alpha(x)\Delta m\,,
\ee
where
\be\begin{split}\label{Delta_m}
\Delta m&=-{q^2\over 2} {\cal G}_\ins{reg}(x,x)=-{q^2\over 2}
\langle\varphi^2\rangle_\ins{ren} \, .
\end{split}\ee
Here ${\cal G}_\ins{reg}(x,x)$ is the coincidence limit $x'\rightarrow x$ of
the regularized Green function
\be
{\cal G}_\ins{reg}(x,x')=
{\cal G}(x,x')-{\cal
G}_\ins{div}(x,x')\, .
\ee
Here the Green function
${\cal G}$ corresponds to the operator  \eq{F}
\be\label{FG}
F\,{\cal G}(x,x')=-
\delta^{D-1}(x,x')\, .
\ee
Thus, in order to find out the self-energy of a scalar charge one has to know
the regularized Euclidean Green function corresponding to the operator \eq{FG}.
In the limit of coincident points this is exactly the
$\langle\varphi^2\rangle_\ins{ren}$ of a free scalar field.  In other words the
problem of calculation of $\Delta m$ is formally equivalent to study of the
quantum fluctuations $\langle\varphi^2\rangle$ in $(D-1)-$dimensional space.
The only difference from our case is that in quantum field theory the
amplitudes of vacuum fluctuations are normalized on $\hbar^{1/2}$ while in the
case of the self-energy of a classical charge they are normalized on unity.
As for calculations it does not make any difference. There are
many well-established methods of calculation of
$\langle\varphi^2\rangle_\ins{ren}$ in quantum field theory. All
traditional methods of UV regularization like point-splitting, zeta-function
and dimensional regularizations, proper time cutoff,  Pauli-Villars and other
approaches are applicable to this task. One can expect that all of them
lead to the same predictions for the $\langle\varphi^2\rangle_\ins{ren}$.

Therefore, technically the problem of calculation of the self-energy of a
charged particle reduces to the calculation of the quantum vacuum average
value of $\langle\varphi^2\rangle_\ins{ren}$.

\section{Calculation of $\boldsymbol{
\langle\varphi^2\rangle_\ins{ren}}$ and its properties}

The Green function \eq{FG} transforms as follows
\be
{\cal G}(x,x')=\Omega^{-{n\over 2}}(x)\,\bar{\cal G}(x,x')\,\Omega^{-{n\over
2}}(x')\,.
\ee
Therefore the non-renormalized value of
$\langle\varphi^2\rangle$ should  transform homogeneously
\be
\langle\varphi^2(x)\rangle=\Omega^{-n}(x)\,\langle\bar{\varphi}^2(x)\rangle\,.
\ee
In other words the combination
\be
g^{n\over 2(n+2)}\langle\varphi^2\rangle
\ee
is an invariant under the transformations \eq{trans}. This classical
symmetry can be broken when one subtracts
divergences from it, therefore
\be
g^{n\over 2(n+2)}\langle\varphi^2\rangle_\ins{ren} \neq \const\,.
\ee
However, one can find such anomalous term $A(x)$ that restores the invariance property
\be\label{phi2}
g^{n\over 2(n+2)}\left(\langle\varphi^2\rangle_\ins{ren}+A\right) = \const\,.
\ee

In order to derive this anomaly we use the Hadamard representation of
the Green function. Consider the divergent part \cite{FrolovZelnikov:2012} of
the Green function
\be\begin{split}\label{calGdiva}
{\cal G}_\ins{div}(x,x')
&={\Delta^{1/2}(x,x'){1\over (2\pi )^{{n\over 2}+1}}}\,\\
&\times\sum_{k=0}^{[n/2]}
{\Gamma\left({{n\over 2}-k}\right)\over
2^{k+1}\sigma^{{n\over2}-k}}{a}_k(x,x') \,.
\end{split}\ee
When $n$ is even the last term $(k=n/2)$ in the sum
should be replaced by
\be\begin{split}
{\Gamma\left({{n\over 2}-k}\right)\over
2^{k+1}{\sigma}^{{n\over
2}-k}}\,&
{a}_k(x,x')\Big|_{k=n/2}\\
&\rightarrow
-{\ln{\sigma}(x,x')+\gamma-\ln
2\over 2^{{n\over
2}+1}}\,{a}_{n/2}(x,x') .
\end{split}\ee
Here ${a}_k(x,x')$  are the Schwinger--DeWitt coefficients
for the operator $F$. The world function
${\sigma}(x,x')$ and Van Vleck--Morette determinant
$\Delta(x,x')$ are defined on the $(n+2)-$dimensional space with the metric $g_{ab}$.
In order to extract the anomaly from the ${\cal G}_\ins{div}$ one has to know how the transformation
\eq{trans} modifies $\sigma,\Delta$, and $a_k$ in the limit of $x'\rightarrow x$.

Here is a list of useful relations
\be\begin{split}
&g_{ab}=\Omega^2\bar{g}_{ab}\hh g^{ab}=\Omega^{-2}\bar{g}^{ab}\,,\\
&g^{ab}\sigma_{a}\sigma_{b}=2\sigma\hh
\bar{g}^{ab}\bar{\sigma}_{a}\bar{\sigma}_{b}=2\bar{\sigma}\,,
\end{split}\ee
where
\be\begin{split}
&\sigma_{a}\equiv\sigma_{,a} \hh \sigma_{ab}\equiv \sigma_{;ab}\,,
\\
&\bar{\sigma}_{a}\equiv\bar{\sigma}_{,a} \hh \bar{\sigma}_{ab}\equiv
\bar{\sigma}_{:ab}\,.
\end{split}\ee
The notation $()_{;}$ means the covariant derivative with respect to the
metric $g_{ab}$, while $()_{:}$ corresponds to the covariant derivative in the
metric $\bar{g}_{ab}$
\be
\sigma_{ab}=g_{ab}-{1\over
3}{\cal R}_{acbd}\,\sigma^c\sigma^d+O(\sigma^{3/2})\,,
\ee
\be
{a}_0(x,x')=1\,,
\ee
\be\begin{split}
\sigma&=\bar{\sigma}\Big[
\Omega^2-\Omega\Omega_{:a}\bar{\sigma}^a\\
&+{1\over12}\left(4\Omega\Omega_{:ab}+4\Omega_{:a}\Omega_{:b}-\bar{g}_
{ ab}\Omega_{:c} \Omega^{:c}
\right)\bar{\sigma}^a\bar{\sigma}^b
\Big]+O(\sigma^{5/2})\,,
\end{split}\ee
\be\begin{split}
\sigma
=\bar{\sigma}\,\Omega(\boldsymbol{x})\Omega(\boldsymbol{x}')
\Big[1+{1\over
12\,\Omega^{2}}(-2\Omega\Omega_{:ab} &\\
+4\Omega_{:a}\Omega_{;b}
-\Omega_{:c}\Omega^{:c}\bar{g}_{ab})\bar{\sigma}^a\bar{\sigma}^b\Big]
&+O(\sigma^{5/2})\,.
\end{split}\ee
For the determinant  $\Delta^{1/2}(\boldsymbol{x},\boldsymbol{x}')$ we
have
\be\begin{split}
\Delta^{1/2}&=1+{1\over
12}{\cal R}_{ab}\sigma^a\sigma^b+O(\sigma^{3/2})\\
&=1+{1\over
12}\bar{{\cal R}}_{ab}\bar{\sigma}^a\bar{\sigma}^b\\
&+{1\over
12\Omega^2}\Big[
-n\Omega\Omega_{:ab}+2n\Omega_{:a}\Omega_{:b}
\\
&-\left(\Omega\Omega_ {:c
}^{:c}+(n-1)\,\Omega^{:c}\Omega_{:c}\right)\bar{g}_{ab}
\Big]\bar{\sigma}^a\bar{\sigma
}^b+O(\sigma^{3/2})\,.
\end{split}\ee

The difference of anomalous terms
\be\label{B_A}
B(x)=A(x)-\Omega^{-n}\bar{A}(x)
\ee
can be defined as
\be
B(x)=\lim_{x'\rightarrow x}\left[ {\cal G}_\ins{div}(x,x')-
{\bar{\cal G}_\ins{div}(x,x')\over\Omega^{{n/
2}}(x)\,\Omega^{{n/
2}}(x')}\right]\,.
\ee

\section{Calculation of a classical self-energy anomaly}

To illustrate the described approach let us consider a couple of
examples.

\subsection{Four dimensions}

In four dimensions $D=4,n=1$
\be\label{Gdiv4}
{\cal
G}_\ins{div}(x,x')
=
{\Delta^{1/2}(x,x')\over 4\pi}\,
{1\over
(2{\sigma})^{1/2}}\,a_0(x,x')\,.
\ee
Thus
\be\begin{split}\label{Gdiv4a}
{\cal G}_\ins{div}(x,x')
&= {1\over 4\pi}\,
{1\over (2{\sigma})^{1/2}}+O(\sigma^{1/2})\\
&= {1\over 4\pi}\,
{1\over
(2{\Omega(x)\bar{\sigma}\Omega(x')})^{1/2}}+O(\sigma^{1/2})\,,\\
\bar{{\cal G}}_\ins{div}(x,x')
&= {1\over 4\pi}\,
{1\over (2{\bar{\sigma}})^{1/2}}+O(\sigma^{1/2})\,,
\end{split}\ee
and, hence,
\be
B(x)=0\,.
\ee
Thus in four dimensions the anomaly vanishes and
\be
\langle\varphi^2\rangle_\ins{ren}=\Omega^{-1}\,\langle\bar{\varphi}
^2\rangle_\ins {ren}\,.
\ee

\subsection{Five dimensions}

In five dimensions $D=5,n=2$
\be\begin{split}\label{Gdiv5}
{\cal
G}_\ins{div}(x,x')
&=
{\Delta^{1/2}(x,x')\over
4\pi^2}\,\left[
{1\over
2{\sigma}}\,{a}_0(x,x')\right.\\
 & \left. - {
1\over
4}(\ln{\sigma}+\gamma-\ln 2)\,{a}_1(x,x')\right] \,.
\end{split}\ee

Taking into account that
\be\begin{split}
{\Delta^{1/2}\over\sigma}&={1\over \Omega\bar{\sigma}\Omega'}\left(
1+{1\over 12}\bar{\cal R}_{ab}\bar{\sigma}^a\bar{\sigma}^b-{1\over 6}\Omega^{-1}\Omega_{:c}^{:c}\,\bar{\sigma}
\right)\,,
\\
{\bar{\Delta}^{1/2}\over\bar{\sigma}}&={1\over\bar{\sigma}}\left(
1+{1\over 12}\bar{\cal R}_{ab}\bar{\sigma}^a\bar{\sigma}^b
\right)\,,
\end{split}\ee
\be
{\cal R}={1\over\Omega^{2}} \left(
    \bar{\cal R} - 6\,\Omega^{-1}\Omega_{:c}^{:c}
    \right)\,,
\ee
\be
V={1\over\Omega^{2}} \left(
    \bar{V}+\,\Omega^{-1}\Omega_{:c}^{:c}
    \right)\,,
\ee
\be\begin{split}
a_1&={1\over 6}{\cal R}+V= {1\over\Omega^{2}}\left({1\over 6}\bar{\cal R}+\bar{V}
    \right)={1\over\Omega^{2}}\,\bar{a}_1\,.
\end{split}\ee
Eventually we obtain
\be\label{B5}
B=-{1\over 48\pi^2}\Omega^{-3}\Omega_{:c}^{:c}-{1\over 8\pi^2}\Omega^{-2}\ln(\Omega)\,\bar{a}_1
\ee
and
\be
\langle\varphi^2\rangle_\ins{ren}=\Omega^{-2}\,\langle\bar{\varphi}
^2\rangle_\ins {ren}-B\,.
\ee

The anomalous contribution $(-B)$ depends both on the physical metric and the
reference one. It would be nice to get such expression
for $A$ which is a functional of only one metric and which gives \eq{B5} after
its substitution to \eq{B_A}. One can easily see that
\be\begin{split}\label{Anomaly}
A(x)&={1\over 288\pi^2}{\cal R}-{1\over 64\pi^2}\ln(g)\,a_1(x)\\
a_1(x)&={1\over 6}{\cal R}+V
\end{split}\ee
is the wanted solution. Though the solution for $A$ may not be uniquely
defined, it is the functional $B$ which is physically important and which is
unambiguous.

%%%%%%%%%%%%%%%%%%%%%%%%

\subsection{Majumdar-Papapetrou spacetimes}

In the paper \cite{FrolovZelnikov:2012a} the self-energy of scalar
charges has been studies in the higher dimensional
Majumdar-Papapetrou spacetime \cite{Myers:1986}. This geometry is the
solution of the Einstein-Maxwell equations which describes
a set of extremely charged black holes in equilibrium in a higher
dimensional asymptotically flat spacetime. The corresponding
background metric and electric potential are
\be\begin{split}\label{MP}
ds^2&=-U^{-2}\,dt^2+U^{2/n}\delta_{ab}\,dx^a dx^b\, ,\\
A_{\mu}&=\sqrt{{n+1\over 2n}}\,U^{-1}\,\delta^0_{\mu}
\hh
n=D-3\,.
\end{split}\ee
Here the function $U$ reads
\be
U=1+\sum_k {M_k\over \rho_k^n}\hh
\rho_k=|\boldsymbol{x}-\boldsymbol{x}_k|\, .
\ee
The index $k=(1,\dots,N)$ enumerates the extremal black holes.
$x^a_k$ is the spatial position of the $k$-th extremal black
hole.

One can see that the transformation \eq{trans} with
\be
\Omega(x) = U^{1/n}(x)
\ee
connects the Majumdar-Papapetrou metric \eq{MP} to the Minkowski
$D-$dimensional metric, that is $(D-1)-$dimensional flat space $\bar{g}_{ab}$
and simultaneously $\bar{\alpha}=1$.
Because in flat spacetime with $\bar{V}=0$ the quantity
$\langle\bar{\varphi}^2\rangle_\ins{ren}=0$ the invariance
property \eq{phi2} automatically gives
\be
\langle\varphi^2\rangle_\ins{ren}=-B\,.
\ee
Then the renormalized self-energy of a scalar charge can be obtained
from Eqs.(\ref{E_ren})-(\ref{Delta_m}).

In four dimensions our approach trivially gives
\be
\Delta m=0\,.
\ee

In five dimensional Majumdar-Papapetrou spacetime, $a_1=0$ and we
reproduce the result \cite{FrolovZelnikov:2012a}
\be
\Delta m={q^2\over 576\pi^2}{\cal R}\, .
\ee
where ${\cal R}$ is the Ricci scalar  of the spatial metric $g_{ab}$.

The analysis of the structure of divergent terms of the scalar Green
function leads to the conclusion that the anomaly in question should
vanish in all even-dimensional spacetimes.

%%%%%%%%%%%%%%%%%%%%%%%%%

\section{Conclusions}

In this paper we have presented an approach to study the self-energy of
pointlike charges based on calculation of the self-energy anomaly. Our approach
is applicable to arbitrary static spacetimes.
The self-energy of static scalar sources of a minimally coupled massless scalar
field is invariant under special symmetry transformations \eq{trans}. These
local transformations consist of simultaneous multiplication of the $g_{tt}$
component
of the metric by a scalar function of spatial coordinates and conformal
transformation of the spatial part $g_{ab}$ of the spacetime metric with the
conformal factor being some power of the same function. In the
case of Majumdar-Papapetrou spacetimes it happens that this symmetry relates
Majumdar-Papapetrou spacetimes to the flat Minkowski spacetime. Therefore the
calculation of the self-energy can be reduced to its calculation in a flat
metric, that
is trivial. The subtle point is that regularization of the UV
divergent quantities violates the invariance property of the classical energy of
the charge distribution and leads to an anomalous
contribution to the renormalized self-energy. In Majumdar-Papapetrou
spacetimes the exact Green functions are known
\cite{FrolovZelnikov:2012} and straightforward calculations
\cite{FrolovZelnikov:2012a} showed that
this anomaly, in fact, constitutes the whole effect. This exact transformation
law makes possible to relate the
self-energy of a charge in the physical spacetime to the self-energy in some
reference spacetime, where its calculation may be significantly simpler. The
proposed approach may provide one with tools for construction of approximate
methods of calculation of $\Delta m$, e.g., similar to the Page approximation
\cite{Page:1982}
developed for calculation of quantum vacuum fluctuations
$\langle\varPhi^2\rangle_\ins{ren}$ and of the stress-energy tensor of fields in
static spacetimes.

\acknowledgments

%\section{Acknowledgments}
This work was partly supported  by  the Natural Sciences and Engineering
Research Council of Canada. The authors are also grateful to the
Killam Trust for its financial support.

\end{document}